# NON-LOCAL MODEL OF HOLLOW CATHODE AND GLOW DISCHARGE – THEORY CALCULATIONS AND EXPERIMENT COMPARISON

Vladimir V. Gorin

*Kyiv National Taras Shevchenko University,
radiophysical faculty,
address: 2 / 5 Glushkov Ave., Kiev 03022, Ukraine*
vgorin@univ.kiev.ua

The non-local equation (V. V. Gorin, 2008 [1 - 3]) for an ionization source

$$s(\mathbf{r}) = \oint_{\partial\Omega} d^2 r' \int d^3 v' G(\mathbf{r},\mathbf{r}',\mathbf{v}') j_n(\mathbf{r}',\mathbf{v}') + \int_\Omega d^3 r' G(\mathbf{r},\mathbf{r}',0) s(\mathbf{r}'), \quad (1)$$

is a fundamental equation in a hollow cathode theory, which allows to formulate a *complete* set of field equations for a self-consistent problem in a stationary glow discharge and a hollow cathode. It enables to describe adequately the region of negative glow and the hollow cathode effect.

Here $s(\mathbf{r})$ is a number of electron and positive ion couples, which are produced in volume unit per time unit as a result of impact ionization, - an *ionization source density*; $\Omega$ is a 3D space region (may be, multiply connected), in which a problem about glow discharge is to be formulated; $\partial\Omega$ is a 2D boundary of the region $\Omega$ (which possibly consists of separate sets: a cathode, an anode, walls of a discharge chamber); $j_n(\mathbf{r},\mathbf{v})d^3v$ is a component of electron flow density, which is normal to the boundary, - the number of electrons, which are emitted from unit area in unit time, their velocity vector in 3D velocity space belongs to interval $d^3v$. The kernel function of the equation is

$$G(\mathbf{r},\mathbf{r}',\mathbf{v}') = \int d^3 v \, Nv\sigma_{ion}(v) g(\mathbf{r},\mathbf{v};\mathbf{r}',\mathbf{v}'), \quad (2)$$

where $N$ is a gas density, $\sigma_{ion}$ is a cross-section of the electron impact ionization, $g(\mathbf{r},\mathbf{v};\mathbf{r}',\mathbf{v}')$ is a source function of the stationary Boltzmann equation for fast (ionizing) electrons:

$$\mathbf{v}\frac{\partial g}{\partial \mathbf{r}} - \frac{e}{m_e}\mathbf{E}(\mathbf{r})\frac{\partial g}{\partial \mathbf{v}} - L(\mathbf{v})g = \delta^3(\mathbf{r}-\mathbf{r}')\delta^3(\mathbf{v}-\mathbf{v}'). \quad (3)$$

The operator $L(\mathbf{v})$ is an arbitrary linear integro-differential operator (with regard to variable $\mathbf{v}$) of electron scattering on neutral atoms or molecules of gas both in elastic and inelastic processes, the choice of which is dependent on a problem complexity (and a possibility to solve it). First term in the right side of the equation (1) at positive value of the flow density $j_n$ gives a source of ionization with electrons, which are emitted from the cathode only; negative values of $j_n$ correspond to the anode and walls, they serve for a source and drain balance of particles, and can be found in result of problem solution. Second term in the right side of the equation describes ionization with secondary electrons, which are generated as a result of impact ionization and have gained energy enough in a region of strong field.

A mathematical kind of the equation (1) in general form is a Fredholm 2-nd type integral equation. In general it is valid for a problem, which has arbitrary geometry configuration of glow discharge, for which a stationary *linear* Boltzmann equation for ionizing electrons is available, that is, whenever neglecting of electron-electron scattering (for ionizing electrons) is possible. Usually a condition of low concentration of plasma in comparison with neutral gas-extender is regarded as sufficient one. In this way slow electrons, which do not have energy enough to produce impact ionization, can be regarded separately, with a Maxwell energy distribution at some temperature (usually, about 1 ev), and can be described as a hydrodynamic fluid (in terms of drift and diffusion). The structure of the equation is universal, but the kernel function $G$ of the equation depends on: 1) phase trajectories of motion of ionizing electrons in concrete electric field configuration, 2) cross-sections of included elastic or inelastic processes, 3) gas-extender density. Together with 1) the Poisson equation, 2) the balance equations for ions and slow electrons, 3) necessary amount of boundary conditions, which include a self-sustained condition, - the equation (1) constitutes a complete set of equations for self-consistent problem of glow discharge or hollow cathode discharge, which must have single solution.

In 1D-configuration, neglecting the elastic scattering and discreteness of energy losses in inelastic processes, the ionizing electrons can be regarded to perform one-dimensional mechanical motion under an accelerating electrical field and a force of friction caused with average energy losses in inelastic processes. The Boltzmann equation allows an analytical solution in this case, and the equation (1) takes a form of the Volterra 2-nd type equation [1, 2].

Let free path of ionizing electrons be small compare with cathode sheath thickness. Then one can neglect inertia of electron and regard its average velocity as established according to local value of the electric field. In this case the kernel of the integral equation degenerates, and the equation (1) can be transformed into the classical Townsend balance equation with local source of ionization: $dj_e/dx = \alpha(E(x))j_e$, which is applied in the Engel and Shteenbeck cathode sheath theory (see [1, 2]}.

It is known that in local model of glow discharge cathode sheath is always followed by positive column. Therefore local model cannot be used both for negative glow simulation and, moreover, for description of hollow cathode discharge. Having comprehension of this and because of absence a formula or an equation for a source of ionization, recent authors of hollow cathode models [4, 5] applied the Monte-Carlo methods for source estimation or attempted to solve many-dimensional Boltzmann equation. The Kolobov and Tsendin paper [6], which aspired to creation «analytic model» of hollow cathode, contains mathematical errors. The very recent paper [7] (2008) of the same Peterburg's school emphasizes an important significance of non-local ionization, but, having no non-local source equation, it offers ungrounded formulae for a source simulation.

Nevertheless, this missing chain in theory is found now.

In present paper an algorithm for calculation of 1D self-consistent problem is created. It includes: 1) the Poisson equation; 2) the balance equation for ions with linear drift and non-local ionization source; 3) the integral equation (1), which has 1D concretizations obtained by V. V. Gorin in paper [1, 2]. Here you can see first attempts to compare calculation results of electrical dependences (pressure - voltage) and experimental data, - under conditions of gradual appearance of the hollow cathode effect. Experimental data for plane hollow cathode with Argon extender, the cathode of which is two disks – bases of the cylinder, the anode is a lateral area of the cylinder, are taken from paper [8]. They are compared with calculations of the theoretical plane hollow cathode, which has «infinite» area of cathode plates. Fig. 1 represents calculation results of the model, in which it is supposed that electron impact ionization occurs *merely* from ground state of Argon, $\varepsilon_{ion} = 15.7$ ev, stepwise ionization through excitation states is not taken into account. In this model the force of friction, $F(w), ev/m$, which affects on ionizing electron in mechanics equations (defining the kernel function in (1)), is regarded to be $F(w) = \varepsilon_{ion} N \sigma_{ion}(w)$, here $N$ is gas density, $\sigma_{ion}$ is ionization cross-section. The dependence of ionization cross-section on the energy is taken as interpolation of the figure in Yu. P. Raizer monograph [9].

In this model the calculated dependences show an appearance of hollow cathode effect – the divergence of lines for *simple glow discharge* (SGD) and *hollow cathode discharge* (HCD) – at some lower pressures than it was in experiment.

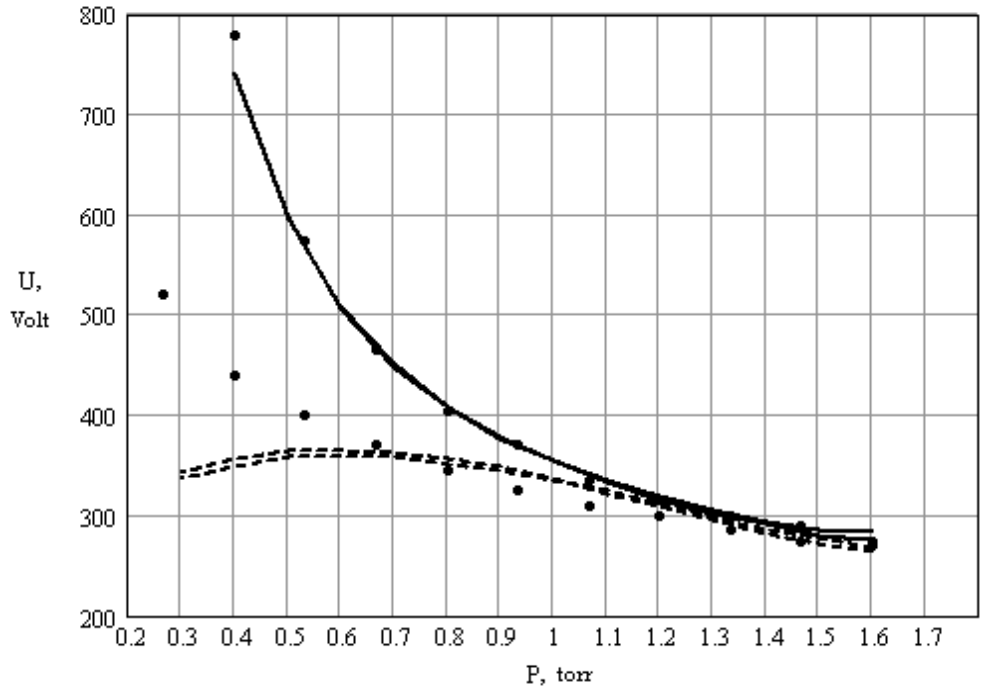

**Fig. 1.** Comparison of theory (lines) and experiment [8] (points). Secondary electron emission factor on the cathode in calculations is γ = 1 / 73. Dash line is a calculation for HCD, solid line – for SGD.
Stepwise ionization is not taken into account.

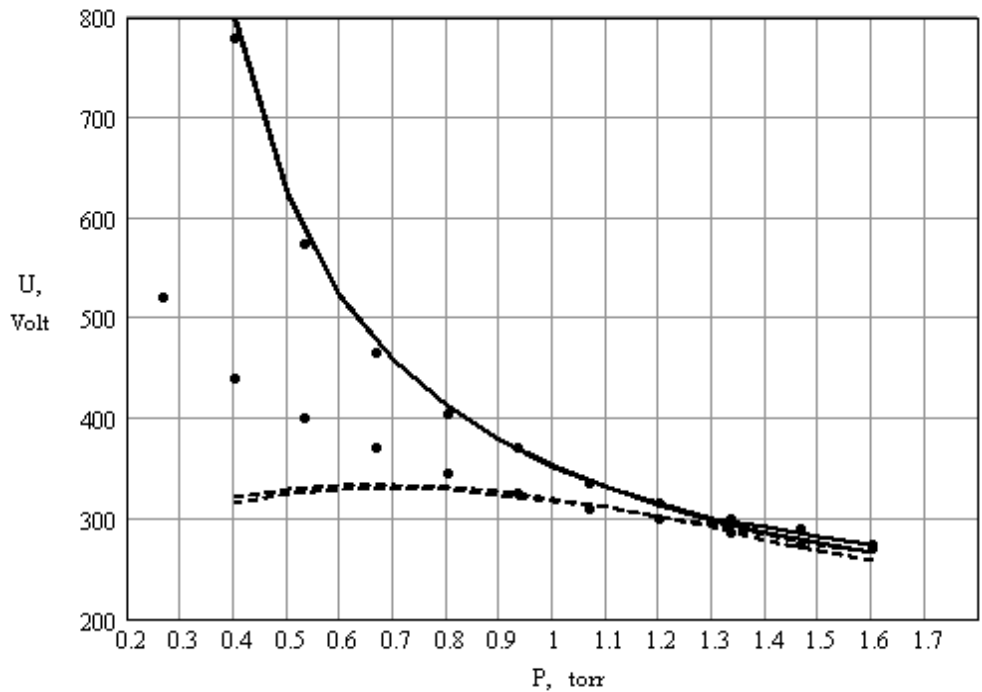

**Fig. 2.** Comparison of theory (lines) and experiment [8] (points). In calculations γ = 1 / 90. Dash line is a calculation for HCD, solid line – for SGD.
Stepwise ionization is taken into account.

At the fig. 2 calculations are made for a model, in which excitation of atoms into first level with energy $\varepsilon_{ex} = 11.5$ ev is regarded as sufficient for atoms to be ionized later with big probability with more slow electrons (because the amount of slow electrons is much greater), - here the mechanism of stepwise ionization is switched on [10]. Energy dependence of excitation cross-sections is simulated with function $\sigma_{ex}(w) \approx \sigma_{ion}(w + \varepsilon_{ion} - \varepsilon_{ex})$ (because of absence the reliable data in the literature about it). The force of friction here is regarded to be $F(w) = \varepsilon_{ex} N \sigma_{ex}(w)$. It is seen from figure 2 that now the calculated hollow cathode effect appears at the same pressures as it was in experiment. The calculated divergence of lines for SGD and HCD is greater than in experiment. It may be caused by finite size of experimental device: space between cathode plates was 2 cm, diameter of plates was 3.14 cm. With fall of pressure the influence of verge effects increases, while in 1D simulation the diameter of plates must be large in comparison with space between them.

It can be seen from pictures, that calculations of theoretical model are in rather good accordance with experimental data. This leads to conclusion, that stepwise ionization in hollow cathode discharge can be significant in the interpretation of current voltage characteristic of the discharge. It is worse to emphasize again, that obtaining of two different curves «pressure-voltage» for simple glow discharge and hollow cathode discharge in frames of *local* classical model is *impossible principally* in itself, because it is impossible to describe pendulum oscillations of electron in neglect of its inertia.